\documentclass[times, 10pt,twocolumn]{article}
\usepackage{latex8}
\usepackage{times}
\usepackage{amsmath,amsthm}
\usepackage{algorithm}
\usepackage{algorithmic}

\newtheorem{definition}{Definition}

\newcommand{\DT}{\emph{\textsf{T}}}
\newcommand{\T}{\textsf{T}}
\newcommand{\DPA}{\textsf{P}}
\newcommand{\PA}{\textsf{P}}
\newcommand{\PRCC}{\textsf{PR\_CC}}
\newcommand{\PRTIP}{\textsf{PR\_TYPE}}
\newcommand{\DPRCC}{\emph{\textsf{PR\_CC}}}
\newcommand{\DPRTIP}{\emph{\textsf{PR\_TYPE}}}
\newcommand{\MPART}{\textsf{MPART}}
\newcommand{\DMPART}{\emph{\textsf{MPART}}}
\newcommand{\MPROT}{\textsf{MPROT}}
\newcommand{\DMPROT}{\emph{\textsf{MPROT}}}
\newcommand{\txtemph}[1]{\text{\emph{#1}}}
\newcommand{\K}{\mathcal{K}}
\newcommand{\TT}{\mathcal{T}}
\newcommand{\DMPARTC}{\emph{\textsf{MPART-C}}}
\newcommand{\DMPROTC}{\emph{\textsf{MPROT-C}}}

\begin{document}

\title{Automated Composition of Security Protocols}

\author{Genge B\'{e}la\\
``Petru Maior'' University of T\^{a}rgu Mure\c{s}\\ Electrical Engineering Department\\
N. Iorga St., No. 1, Tg. Mure\c{s}, Romania\\
bgenge@engineering.upm.ro\\
\and
Iosif Ignat\\
Technical University of Cluj-Napoca\\
Computer Science Department\\
Gh. Baritiu St., No. 28, Cluj-Napoca, Romania\\
Iosif.Ignat@cs.utcluj.ro\\
\and
Haller Piroska\\
``Petru Maior'' University of T\^{a}rgu Mure\c{s}\\ Electrical Engineering Department\\
N. Iorga St., No. 1, Tg. Mure\c{s}, Romania\\
phaller@engineering.upm.ro\\
}

\maketitle \thispagestyle{empty}

\begin{abstract}
Determining if two protocols can be securely composed requires
analyzing not only their additive properties but also their
destructive properties. In this paper we propose a new composition
method for constructing protocols based on existing ones found in
the literature that can be fully automatized. The additive
properties of the composed protocols are ensured by the composition
of protocol preconditions and effects, denoting, respectively, the
conditions that must hold for protocols to be executed and the
conditions that hold after executing the protocols. The
non-destructive property of the final composed protocol is verified
by analyzing the independence of the involved protocols, a method
proposed by the authors in their previous work. The fully
automatized property is ensured by constructing a rich protocol
model that contains explicit description of protocol preconditions,
effects, generated terms and exchanged messages. The proposed method
is validated by composing 17 protocol pairs and by verifying the
correctness of the composed protocols with an existing tool.
\end{abstract}

\section{Introduction}

Security protocols are ``communication protocols dedicated to
achieving security goals'' (C.J.F. Cremers and S. Mauw) \cite{CM05}
such as confidentiality, integrity or availability. Achieving such
security goals is made through the use of cryptography. The
explosive development of today's Internet and the technological
advances made it possible to implement and use security protocols in
a wide range of applications such as sensor networks, electronic
commerce or routing environments.

Security protocols have been intensively analyzed throughout the
last few decades, resulting in a variety of dedicated formal methods
and tools \cite{FHGC99,Wei99,CRE06b}. The majority of these methods
consider a Dolev-Yao-like intruder model \cite{DY83,Cer01} to
capture the actions available to the intruder that has complete
control over the network. By analyzing each individual protocol in
the presence of this intruder, the literature has reported numerous
types of attacks \cite{Wei99,Gl96}. However, in practice, there can
be multiple protocols running over the same network, thus the
intruder is given new opportunities to construct attacks by
combining messages from several protocols, also known as
multi-protocol attacks \cite{Cre06a}.

Designing new protocols, thus, becomes a challenging task if we look
at the number of attacks that have been discovered over the years
\cite{Gl96} after the protocols have been published. In the last few
years the use of protocol composition \cite{Cre06a,Choi06,Can05} has
been successfully applied to create new protocols based on existing
\cite{DDMR03, DDMR07, ACGMMR08} or predefined protocols
\cite{Choi06}.

In this paper we propose a new composition method that, as opposed
to existing approaches \cite{Choi06, DDMR03, DDMR07, ACGMMR08,
Gut02} can be fully automatized by eliminating the human factor. In
order to create an automated composition method, we need an enriched
protocol model that contains enough information to compose the
protocol preconditions and effects and an approach for the
verification of the correctness of the final, composed protocol.

Preconditions denote the set of properties that must be satisfied
for the protocol to be executed, while the effects denote the set of
properties resulting from the protocol execution. By composing
preconditions and effects (i.e. PE composition), we generate a new
protocol sequence that ensures the satisfaction of the protocol
preconditions and the propagation of generated information through
effects.

The protocol sequence generated by the PE composition must be
correct, in the sense that it must maintain the security properties
of the original protocols. In order to verify this, we use an
approach developed in our previous work \cite{GI07} that verifies
the independence of the involved protocols. Protocol independence,
called participant chain composition (i.e. PC composition) ensures that the intruder can not replay
messages from one protocol to another to construct new attacks while
running the protocols in the same environment. This property
also ensures the correctness of the composed protocol.

The paper is structured as follows. In section \ref{SEC:PROTMODEL}
we define an enriched protocol model that includes explicit
description of protocol preconditions, effects, generated terms and
exchanged messages. In section \ref{SEC:COMPOSITION} we provide a
description of the proposed composition method and a brief
presentation of the independence verification method proposed in our
previous work \cite{GI07}. The proposed composition method has been
applied in the composition process of several protocols, part of
these experimental results are given in section
\ref{SEC:EXPRESULTS}. We relate our work to others found in the
literature in section \ref{SEC:RELWORK} and we end with a conclusion
and future work in section \ref{SEC:CONCLUSION}.

\section{Protocol model}\label{SEC:PROTMODEL}

Protocol participants communicate by exchanging \emph{terms}
constructed from elements belonging to the following basic sets:
\textsf{P}, denoting the set of participant names; \textsf{N},
denoting the set of random numbers or \emph{nonces} (i.e. ``number
once used''); \textsf{K}, denoting the set of cryptographic keys;
\textsf{C}, denoting the set of certificates and \textsf{M},
denoting the set of user-defined message components.

In order for the protocol model to capture the message component
types found in security protocol implementations \cite{SAML07,
WSSECURITY06} we specialize the basic sets with the following
subsets:
\begin{itemize}
  \item $\textsf{P}_{DN}\subseteq\textsf{P}$, denoting the set of distinguished names; $\textsf{P}_{UD}\subseteq\textsf{P}$, denoting the set of user-domain names; $\textsf{P}_{IP}\subseteq\textsf{P}$, denoting the set of user-ip names; $\textsf{P}_{U}=\{\textsf{P}\setminus\{\textsf{P}_{DN}\cup\textsf{P}_{UD}\cup\textsf{P}_{IP}\}\}$,
      denoting the set of names that do not belong to the previous subsets;
  \item $\textsf{N}_T$, denoting the set of timestamps;
      $\textsf{N}_{DH}$, denoting the set of random numbers specific to the Diffie-Hellman key exchange;
      $\textsf{N}_A=\{\textsf{N}\setminus\{\textsf{N}_{DH}\cup\textsf{N}_T\}\}$, denoting the set of random numbers;
  \item $\textsf{K}_S\subseteq\textsf{K}$, denoting the set of symmetric keys; $\textsf{K}_{DH}\subseteq\textsf{K}$, denoting the set of keys generated from a Diffie-Hellman key exchange;
      $\textsf{K}_{PUB}\subseteq\textsf{K}$, denoting the set of public keys;
      $\textsf{K}_{PRV}\subseteq\textsf{K}$, denoting the set of private keys.
\end{itemize}

To denote the encryption type used to create cryptographic terms, we
define the following \emph{function names}:
\begin{align*}
FuncName & ::=sk && (symmetric\,function)\\
&\quad\mid pk         && (asymmetric\,function)\\
&\quad\mid h          &&  (hash\,function)\\
&\quad\mid hmac       &&  (keyed\,hash\,function)
\end{align*}

The encryption and decryption process makes use of cryptographic
keys. Decrypting an encrypted term is only possible if participants
are in the possession of the decryption key pair. In case of
symmetric cryptography, the decryption key is the same as the
encryption key. In case of asymmetric cryptography, there is a
public-private key pair. Determining the corresponding key pair is
done using the function $\_^{-1}:\textsf{K}\rightarrow\textsf{K}$.

The above-defined basic sets and function names are used in the
definition of \emph{terms}, where we also introduce constructors for
pairing and encryption:
\begin{align*}
\textsf{T}::=.\mid \textsf{R} \mid \textsf{N} \mid \textsf{K} \mid
\textsf{C} \mid \textsf{M} \mid (\textsf{T},\textsf{T}) \mid
\{\textsf{T}\}_{FuncName(\textsf{T})},
\end{align*}
where the `.' symbol is used to denote an empty term.

Having defined the terms exchanged by participants, we can proceed
with the definition of a \emph{node} and a \emph{participant chain}.
To capture the sending and receiving of terms, the definition of
nodes uses \emph{signed terms}. The occurrence of a term with a
positive sign denotes transmission, while the occurrence of a term
with a negative sign denotes reception.

\begin{definition}
A \emph{node} is any transmission or reception of a term denoted as
$\langle\sigma,t\rangle$, with $t\in\DT$ and $\sigma$ one of the
symbols $+,-$. A node is written as $-t$ or $+t$. We use $(\pm\DT)$
to denote a set of nodes. Let $n\in(\pm\DT)$, then we define the
function $sign(n)$ to map the sign and the function $term(n)$ to map
the term corresponding to a given node.
\end{definition}

\begin{definition}
A \emph{participant chain} is a sequence of nodes. We use
$(\pm\DT)^\ast$ to denote the set of finite sequences of nodes and
$\langle \pm t_1, \pm t_2, \ldots, \pm t_i \rangle$ to denote an
element of $(\pm\DT)^\ast$.
\end{definition}

In order to define a participant model we also need to define the
preconditions that must be met such that a participant is able to
execute a given protocol. In addition, we also need to define the
effects resulting from a participant executing a protocol.

Preconditions and effects are defined using predicates applied on
terms: $\txtemph{CON\_TERM}:\T$, denoting a term that must be
previously generated (preconditions) or it is generated (effects);
$\txtemph{CON\_PARTAUTH}:\T$, denoting a participant that must be
previously authenticated (preconditions) or a participant that is
authenticated (effects); $\txtemph{CON\_CONF}:\T$, denoting that a
given term must be confidential (preconditions) or it is kept
confidential (effects); $\txtemph{CON\_INTEG}:\T$, denoting that for
a given term the integrity property must be provided (preconditions)
or that the protocol ensures the integrity property for the given
term (effects); $\txtemph{CON\_NONREP}:\T$, denoting that for a
given term the non-repudiation property must be provided
(preconditions) or that the protocol ensures the non-repudiation
property for the given term (effects); $\txtemph{CON\_KEYEX}:\T$,
denoting that a key exchange protocol must be executed before
(preconditions) or that this protocol provides a key exchange
resulting the given term (effects).

The set of precondition-effect predicates is denoted by $\PRCC$ and
the set of precondition-effect predicate subsets is denoted by
$\PRCC^*$. Next, we define predicates for each type of term
exchanged by protocol participants. These predicates are based on
the basic and specialized sets provided at the beginning of this
section. We use the $\txtemph{TYPE\_DN}:\T$ predicate to denote
distinguished name terms, $\txtemph{TYPE\_UD}:\T$ to denote
user-domain name terms, $TYPE\_IP:\T$ to denote user-ip name terms,
$\txtemph{TYPE\_U}:\T$ user name terms, $\txtemph{TYPE\_NT}:\T$ to
denote timestamp terms, $\txtemph{TYPE\_NDH}:\T$ to denote
Diffie-Hellman random number terms, $\txtemph{TYPE\_NA}:\T$ to
denote other random number terms,
$\txtemph{TYPE\_NDH}:\T\times\T\times\T\times\PA\times\PA$ to denote
Diffie-Hellman symmetric key terms $(term, number_1, number_2,$
$participant_1,participant_2)$,
$\txtemph{TYPE\_KSYM}:\T\times\PA\times\PA$ to denote symmetric key
terms $(term, participant_1, participant_2)$,
$\txtemph{TYPE\_KPUB}:\T\times\PA$ to denote public key terms
$(term, participant)$, $\txtemph{TYPE\_KPRV}:\T\times\PA$ to denote
private key terms $(term, participant)$,
$\txtemph{TYPE\_CERT}:\T\times\PA$ do denote certificate terms
$(term, participant)$ and $\txtemph{TYPE\_MSG}:\T$ to denote
user-defined terms.

The set of type predicates is denoted by $\PRTIP$ and the set of
type predicate subsets is denoted by $\PRTIP^*$. Based on the
defined sets and predicates we are now ready to define the
participant and protocol models.

\begin{definition}
A \emph{participant model} is a tuple $\langle
prec,\txtemph{\emph{eff}},type,gen,part,chain\rangle$, where
$prec\in\DPRCC^*$ is a set of precondition predicates,
$\txtemph{\emph{eff}}\in\DPRCC^*$ is a set of effect predicates,
$type\in\DPRTIP$ is a set of type predicates, $gen\in\DT^*$ is a set
of generated terms, $part\in\DPA$ is a participant name and
$chain\in(\pm\DT)^*$ is a participant chain. We use the $\DMPART$
symbol to denote the set of all participant models.
\end{definition}

\begin{definition}
A \emph{protocol model} is a collection of participant models such
that for each positive node $n_1$ there is exactly one negative node
$n_2$ with $term(n_1)=term(n_2)$. We use the $\DMPROT$ symbol to
denote the set of all protocol models.
\end{definition}

\section{Composition of protocol models}
\label{SEC:COMPOSITION}

The composition process involves composing in a first stage the
protocol preconditions and effects followed by the composition of
participant chains. In this section we first formulate the
conditions needed for the precondition-effect (PE) composition which
involves establishing the satisfaction of protocol preconditions and
the verification of the non-destructive properties of protocol
effects. This is followed by the protocol-chain (PC) composition for
which we construct a canonical model and verify the independence of
the involved participant chains.

\subsection{Composition of preconditions and effects}

In the composition process of two security protocols we first need
to compose the preconditions and effects. In other words, we need to
establish if the knowledge needed by protocol participants to run a
given protocol, expressed through the form of precondition
predicates, is available and if the set of precondition and effect
predicates is non-destructive.

In order to establish if the set of preconditions corresponding to a
protocol can be satisfied based on a given context and the effects
corresponding to another protocol we use the predicate
$\txtemph{PART\_PREC}:\txtemph{\DT}^*\times\PRCC^*\times\PRCC^*$.
The context denotes the initial knowledge available to participants
when running the protocol. For two participant models,
$\varsigma_1=\langle
prec_1,\txtemph{eff}_1,type_1,gen_1,part_1,chain_1\rangle$ and
$\varsigma_2=\langle
prec_2,\txtemph{eff}_2,type_2,gen_2,part_2,chain_2\rangle$, the
$\txtemph{PART\_PREC}$ predicate is defined as
$$
\begin{array}{l}
\setlength\arraycolsep{2pt} \txtemph{PART\_PREC}(ctx,\txtemph{eff}_1,prec_2) = \\
\left\lbrace
  \begin{array}{ll}
    True,      &\text{ if $\emph{eff}_1\subseteq prec_2\cup$ },\\
               &\text{ $\quad\{\cup\{CON\_TERM(t)|t \in ctx\}\}$ },\\
    False,     &\text{ otherwise }.
  \end{array}
\right.
\end{array}
$$

The non-destructive property applies only for the
$\txtemph{CON\_CONF}$ because the absence of another property, such
as integrity or non-repudiation, does not affect the previous
properties. In order to establish if the preconditions and effects
of two participant models are destructive we use the predicate
$\txtemph{PART\_NONDESTR}:\PRCC^*\times\PRCC^*\times\PRCC^*$ which
holds only if all confidential terms from one participant model
maintain their confidentiality property in the second participant
model also. Thus, the predicate is defined as
$$
\begin{array}{l}
\setlength\arraycolsep{2pt}
 \txtemph{PART\_NONDESTR}(\txtemph{eff}_1,prec_2,\txtemph{eff}_2) =\\
 \left\lbrace
   \begin{array}{ll}
    True,      &\text{if $EF_1\neq\txtemph{CON\_CONF}\vee$}\\
               &\text{\qquad if $EF_1=\txtemph{CON\_CONF}\wedge t_1=t_2$ then}\\
               &\text{\qquad\quad$\exists EF_2(t_2):EF_2=\txtemph{CON\_CONF}$,}\\
               &\text{$ \forall EF_1(t_1)\in \emph{eff}_1 \wedge \forall PR_2(t_2)\in prec_2$,}\\
    False,     &\text{otherwise}.
  \end{array}
\right.
\end{array}
$$

Based on the above given predicates we can state that in order to
compose the preconditions and effects corresponding to two
participant models we need to establish if the predicates
$\txtemph{PART\_PREC}$ and $\txtemph{PART\_NONDESTR}$ hold. The
precondition-effect (PE) composition is expressed through the use of
the operator
$\_\prec_{\varsigma}^{PE}\_:\MPART\times\MPART\rightarrow\MPART$,
which generates a new participant model based on two given
participant models. By using this operator, we not only express the
PE composition of participant models but also the order in which the
given participant models appear in the final, composed participant
model. Thus, we can state that given two participant models,
$\varsigma_1$ and $\varsigma_2$, for which the PE composition
requirements are satisfied, we have that
$\varsigma_1\prec_{\varsigma}^{PE}\varsigma_2\neq\varsigma_2\prec_{\varsigma}^{PE}\varsigma_1$.
If the operator is applied on two participant models that can not be
composed (i.e. one of the two predicates does not hold), the result
is the empty participant model, denoted by $\phi_{\varsigma}=\langle
\phi, phi, \phi, \phi, ., \langle\rangle \rangle$, where $\phi$
denotes an empty set.

The PE composition requirements of two participant models can be
easily extended to form the requirements for the PE composition of
two protocol models. These requirements include applying the
$\_\prec_{\varsigma}^{PE}\_$ operator on pairs of participant models
for which the names are equal. We express the PE composition of two
protocol models through the use of the
$\_\prec_{\xi}^{PE}\_:\MPROT\times\MPROT\rightarrow\MPROT$ operator.
For this operator also, we can state that given two protocol models,
$\xi_1$ and $\xi_2$, for which the PE composition requirements are
satisfied, we have that
$\xi_1\prec_{\xi}^{PE}\xi_2\neq\xi_2\prec_{\xi}^{PE}\xi_1$. In case
of protocol models that can not be composed, the result is denoted
by the empty protocol model $\phi_{\xi}=\phi$.

\subsection{Composition of participant chains}

The PC composition makes use of a canonical model that focuses on
terms that can be verified by protocol participants. For each term
the canonical model provides a corresponding syntactical
representation through the use of \emph{basic types}. These denote
the terms that can be verified by protocol participants also
including a representation for terms that can not be verified
because of limited participant knowledge. The verification process
makes use of these types to decide if attacks can be constructed on
each protocol model by using terms extracted from the other
considered protocol models.

In order to compose two participant chains these must be
\emph{instance independent} and \emph{canonical independent}. The
first condition refers to the non-destructive properties of
preconditions and effects while the second condition refers to
verifying the independence of the involved participant chains based
on the canonical model. The verification of the independence
property of protocol models has been covered by the authors in their
previous work \cite{GI07}. If protocols are independent, then they
maintain their security properties when they are run in the same
context. By using this property in the composition process,
protocols maintain their security properties, resulting new
protocols with accumulated properties.

In the remaining of this section we briefly present the canonical
model and the protocol independence property proposed in our
previous work.

The \emph{basic types} we consider are based on the specialized
basic sets introduced in the protocol model:
$$
\begin{array}{ll}
BasicType::= & \textsf{p}_{DN} \mid \textsf{p}_{UD} \mid \textsf{p}_{IP} \mid \textsf{p}_{U} \mid \textsf{n}_{T} \mid \textsf{n}_{DH} \\
&\mid \textsf{n}_{A} \mid \K \mid \textsf{m} \mid \textsf{c} \mid
\textsf{u},
\end{array}
$$
where the given symbols correspond to participant distinguished
names, user-domain names, user-ip names, other user names,
timestamps, Diffie-Hellman random numbers, other random numbers,
keys, user defined terms, certificates and unknown terms,
respectively.

The \emph{unknown} type \textsf{u} corresponds to terms that can not
be validated because of limited participant knowledge. By including
this information in the specification we are able to detect subtle
type-flaw attacks using a syntactical comparison of typed terms,
that otherwise would require the construction of a state-space that
can become rather large if we consider the existence of multiple
protocols in the same system \cite{Cre05}.

Based on the defined basic terms we can now proceed with the
definition of \emph{canonical terms}:
\begin{align*}
\TT::=.\mid BasicType \mid (\TT,\TT) \mid \{\TT\}_{FuncName(\TT)}.
\end{align*}

A canonical node is defined as a signed canonical term using the
following definition.

\begin{definition}
A \emph{canonical node} is any transmission or reception of a
canonical term denoted by $\langle\sigma,t\rangle$, with $t\in\TT$
and $\sigma$ one of the symbols $+,-$. We use $(\pm\TT)$ to denote a
set of canonical nodes. Let $n\in(\pm\TT)$, then we define the
function $csign(n)$ to map the sign and the function $cterm(n)$ to
map the canonical term corresponding to a given canonical node.
\end{definition}

Before we proceed with the definition of canonical chains and
canonical participant models we need to define \emph{classifiers}.
These are attached to participant chains and are used to transform
canonical terms received from other participants based on local
participant knowledge. We define two such classifiers:
\begin{align*}
Classifier::= CL_P \mid CL_V.
\end{align*}

The first classifier $CL_P$ denotes the processing chain
corresponding to a participant. This chain contains canonical terms
that correspond to participant knowledge. The second classifier
$CL_V$ denotes the virtual chain used to transform received terms
from the transmitted form to the received form based on the
knowledge of the receiving participant.

\begin{definition}
A \emph{canonical participant chain} is a sequence of canonical
nodes. A \emph{classified canonical participant chain} is a pair
$\langle CL, l_{cc}\rangle$, where $CL\in Classifier$ and $l_{cc}\in
(\pm\TT)^*$. We use $(\pm\TT)^*$ to denote a set of canonical
participant chains.
\end{definition}

\begin{definition}
A \emph{canonical participant model} is a pair $\langle
part,sl_{cc}\rangle$, where $part\in\PA$ is a participant name and
$sl_{cc}\in(Classifier\times(\pm\TT)^*)^*$ is a set of classified
canonical participant chains. We use $\DMPARTC$ to denote the set of
all canonical participant models.
\end{definition}

Next, we define a canonical protocol model as a set of canonical
participant models.

\begin{definition}
A \emph{canonical protocol model} is a collection of canonical
participant models such that for each positive canonical node $n_1$
there is exactly one negative canonical node $n_2$ with
$cterm(n_1)=cterm(n_2)$. We use the $\DMPROTC$ symbol to denote the
set of all canonical protocol models.
\end{definition}

Based on the described protocol and canonical models, we proved,
through the form of a proposition, that if two protocol models are
instance independent and their corresponding canonical models are
canonical independent, then the intruder can not construct attacks
using terms extracted from other protocols. In order to verify this
we used an intruder model based on the Dolev-Yao \cite{DY83,Cer01}
model to capture the powers that can be used by an intruder.

If two protocol models are independent, then their participant
chains can be composed. We use the
$\_\prec_{\varsigma}^{PC}\_:\MPART\times\MPART\rightarrow\MPART$
operator to denote the PC composition of protocol chains and the
$\_\prec_{\xi}^{PC}\_:\MPROT\times\MPROT\rightarrow\MPROT$ operator
to denote the PC composition of protocol models. For the first
operator we use $\phi_{\varsigma}$ to denote the empty participant
model, while for the second operator we use $\phi_{\xi}$ to denote
the empty protocol model.

If two protocol models can be composed PE and PC, then they can be
composed. The composition operator we use to denote the composition
of protocol models is
$\_\prec^{C}\_:\MPROT\times\MPROT\rightarrow\MPROT$, for which the
generated empty protocol model is denoted by $\phi_{\xi}$.

By sequentially composing several protocol models the resulting
protocol model provides a unified set of preconditions and effects
and a unified set of participant chains. By composing \emph{i}
protocols, the resulting sequence is written as
$\xi_1\prec^{C}\xi_2\prec^{C}\ldots\prec^{C}\xi_i$.

\subsection{Composition algorithm}

The proposed composition method can be applied on protocol pairs or
entire protocol sequences. Let $SEQ_1$ and $SEQ_2$ be two protocol
sequences, where each sequence is constructed by subsequently
applying the $\_\prec^{C}\_$ operator on protocol pairs, and $n$,
$m$, two symbols denoting the number of protocols in the first and
in the second sequence, respectively. Then, the composition
algorithm must ensure that the new composed sequence maintains the
security properties of the original protocols and that the knowledge
available to protocol participants allows the execution of the new
sequence.
\begin{algorithm}
\caption{Composition steps} \label{ALG:COMP}
\begin{algorithmic}
\STATE \COMMENT{Verification of non-destructive properties} \FORALL
{$\xi_1\in SEQ_1$ and $\xi_2 \in SEQ_2$}
    \FORALL {$\varsigma_1\in\xi_1$ and $\varsigma_2\in\xi_2$}
    \STATE Let $\varsigma_1=\langle prec_1,\txtemph{eff}_1,type_1,gen_1,part_1,chain_1\rangle$,
    \STATE $\phantom{Let }\,\varsigma_2=\langle prec_2,\txtemph{eff}_2,type_2,gen_2,part_2,chain_2\rangle$,
    \STATE $\phantom{Let }\,c_1 =\txtemph{PART\_NONDESTR}(\txtemph{eff}_1,prec_2,\txtemph{eff}_2)$,
    \STATE $\phantom{Let }\,c_2 = \txtemph{PART\_NONDESTR}(\txtemph{eff}_2,prec_1,\txtemph{eff}_1)$
    \IF {$c_1=False\vee c_2=False\vee\varsigma_1\prec_{\varsigma}^{PC}\varsigma_2=\phi_{\varsigma}$}
    \STATE $@InterruptExecution$
    \ENDIF
    \ENDFOR
\ENDFOR \STATE \COMMENT{Composition of protocol sequences} \STATE
Let $i = 1, j = 1$ \STATE Let $\xi=\{\langle \phi, \txtemph{PRINIT},
\txtemph{TINIT}, \phi, ., \phi \rangle\}$ \WHILE {$i \leq n \wedge j
\leq m$}
    \STATE Let $\xi^i$ be the $i$-th element of $SEQ_1$
    \STATE Let $\xi^j$ be the $j$-th element of $SEQ_2$
    \IF {$\xi\prec_{\xi}^C\xi^i\neq\phi_{\xi}$}
    \STATE $\xi=\xi^i\prec_{\xi}^C\xi$, $i = i + 1$
    \ELSIF {$\xi^i\prec_{\xi}^C\xi\neq\phi_{\xi}$}
    \STATE $\xi=\xi\prec_{\xi}^C\xi^i$, $i = i + 1$
    \ENDIF
    \IF {$\xi\prec_{\xi}^C\xi^j\neq\phi_{\xi}$}
    \STATE $\xi=\xi^j\prec_{\xi}^C\xi$, $j = j + 1$
    \ELSIF {$\xi^j\prec_{\xi}^C\xi\neq\phi_{\xi}$}
    \STATE $\xi=\xi\prec_{\xi}^C\xi^j$, $j = j + 1$
    \ENDIF
\ENDWHILE \STATE \COMMENT{Add remaining protocols} \WHILE {$i \leq
n$} \STATE $\xi=\xi\prec_{\xi}^C\xi^i$, $i = i + 1$ \ENDWHILE \WHILE
{$j \leq m$} \STATE $\xi=\xi\prec_{\xi}^C\xi^j$, $j = j + 1$
\ENDWHILE
\end{algorithmic}
\end{algorithm}
Verifying if protocols from the two sequences maintain their
security properties requires applying the $\txtemph{PART\_NONDESTR}$
predicate on each protocol pair and the verification of the
independence of the participant chains by using the PC composition
operator $\_\prec_{\varsigma}^{PC}\_$. As shown in Algorithm
\ref{ALG:COMP}, if one of these conditions is not satisfied, the
execution is stopped, symbolized using the $@InterruptExecution$
keyword.

If the protocol properties are not destructive, the execution of the
composition algorithm continues with the composition of protocol
components. The final protocol is denoted by $\xi$, which,
initially, contains a participant model with the effects
$\txtemph{PRINIT}$ and types $\txtemph{TINIT}$. These denote the
initial knowledge for protocol participants, extracted from the
context $ctx$, a unified context constructed from the contexts
corresponding the the two sequences.

The composition process locates the position of each protocol in the
final sequence by using the composition operator
$\_\prec_{\varsigma}^C\_$. If the result is $\phi_{\varsigma}$, the
protocols can not be composed and another pair is selected. Finally,
the remaining protocols are added to the sequence.

\section{Experimental results}\label{SEC:EXPRESULTS}

In order to validate the proposed method we generated several new
composed protocols, based on existing ones. In order to verify if
the new protocols accumulated the properties of the initial
protocols, i.e. the composition is non-destructive, we applied the
method proposed in this paper. However, such a verification is not
enough for validating a method that must ensure the correctness of
the resulted protocols, as shown by the large number of attacks
discovered on protocols long after they have been published
\cite{Wei99, Gl96}.

Having these in mind, we turned to existing protocol verification
tools. The purpose of the verification was to determine if new
attacks became available on the composed protocols. One of the few
tools allowing the verification of multi-protocol attacks is Scyther
\cite{CRE06b}, which is the only tool currently available that also
detects type-flaw attacks \cite{HLS00, Mea03}, commonly found in
multi-protocol environments.

We have applied our method on several pairs of security protocols
defined in the library maintained by Clark and Jacob \cite{CJ97},
for which there is also an online version available \cite{SPORE08}.
Through our experiments we composed protocol pairs such as CCITT
X.509 v1 (i.e. X509v1) and CCITT X.509 v1c (i.e. X509v1c), BAN
Concrete RPC (i.e. BAN-RPC) and Lowe-B (i.e. Lowe-BAN),
Lowe-Denning-Sacco (i.e. L-D-S) and Kao-Chow v1 (i.e. K-Cv1),
Lowe-Kerberos (i.e. Lowe-Kerb) and Neuman-Stubblebine (i.e.
Neuman-S), Hwang-Neuman-Stubblebine (i.e. H-N-S) and
Neuman-Stubblebine, Needham-Schroeder (i.e. Needh-S) and CCITT X.509
v1, Lowe-Needham-Schroeder (i.e. L-N-S) and ISO9798, Otway-Rees
(i.e. Otway-R) and Lowe-BAN, Yahalom-Lowe (i.e. Y-L) and Kao-Chow
v1, as shown in Table \ref{TAB:EXPRESULTS}. The non-destructive
property of the composed protocol was validated using the Scyther
tool.

In Table \ref{TAB:EXPRESULTS}, S1 indicates the protocol composition sequence P1-P2, while S2 indicates the sequence P2-P1. We used ``Y''
to indicate the successful composition of a sequence and ``N'' the
failure of the composition process. By applying the proposed
non-destructivity conditions we have discovered several new
multi-protocol attacks. For example, in case of the protocol pair
Yahalom-Lowe and Kao-Chow, we discovered a new attack that gives the
intruder the possibility to replay valid messages from the Kao-Chow
v1 (i.e. K-Cv1) protocol into the Yahalom-Lowe (i.e. Y-L) protocol.
We have created a composed protocol and used the Scyther tool to
verify it. The result was that 2 new attacks were possible. After
correcting the problem by adding additional terms to the protocols
messages in order for participants to be able to verify the validity
of these messages, the Scyther tool did not detect any attacks,
which was also confirmed by our method.
\begin{table}
\centering \caption{Protocol composition results}
\label{TAB:EXPRESULTS}       
\begin{tabular}{|l|l|c|c|c|}
\hline
\textbf{Protocol 1}    &  \textbf{Protocol 2} & \textbf{PE} & \textbf{PC}    & \textbf{Scyther} \\
& & \textbf{(S1/S2)}& \textbf{(S1/S2)} &  \\
\hline\hline
Lowe-B    &  ISO9798  & N/Y    & Y/Y   & Y/Y \\
\hline
Lowe-B    &  X509v1  & N/N     & Y/Y   & Y/Y \\
\hline
ISO9798     &  X509v1  & Y/Y   & Y/Y   & Y/Y \\
\hline
ISO9798     &  X509v1c & Y/Y   & Y/Y   & Y/Y \\
\hline
X509v1     &  X509v1c & Y/Y   & Y/Y   & Y/Y \\
\hline
X509v1     &  X509v1c & Y/Y   & Y/Y   & Y/Y \\
\hline
BAN-RPC    & Lowe-B  & Y/Y   & N/N     & N/N \\
\hline
L-D-S       & K-Cv1     & Y/Y   & N/N     & N/N \\
\hline
K-Cv1       & K-Cv2     & Y/Y   & Y/Y   & Y/Y \\
\hline
L-D-S       & Kerbv5& Y/Y   & N/N     & N/N \\
\hline
Lowe-Kerb   & Neuman-S  & Y/Y   & N/N     & N/N \\
\hline
H-N-S       & Neuman-S  & Y/Y   & Y/Y   & Y/Y \\
\hline
Needh-S   & X509v1    & Y/N    & Y/Y   & Y/Y \\
\hline
L-N-S       & ISO9798   & Y/N    & Y/Y   & Y/Y \\
\hline
Otway-R     & Lowe-B  & Y/N    & Y/Y   & Y/Y \\
\hline
SPLICE      & Needh-S & Y/Y   & Y/Y   & Y/Y \\
\hline
TMN         & Andr-RPC& Y/N    & Y/Y   & Y/Y \\
\hline
Y-L         & K-Cv1     & Y/Y   & N/N     & N/N \\
\hline
\end{tabular}
\end{table}
\section{Related work}\label{SEC:RELWORK}

In this section we briefly describe the approaches found in
literature that mostly relate to our proposal.

In \cite{Gut02}, Guttman proposes a composition method based on
predefined protocol primitives that are used to construct new,
composed protocols. A similar approach is proposed by Choi
\cite{Choi06}, that additionally defines \emph{bindings} in order to
correctly connect different primitives. The previously mentioned
approaches have not been designed to compose existing protocols, as
the one proposed in this paper. We have only mentioned them here for
completeness.

A. Datta et all \cite{DDMR03, DDMR07} propose the description of
each composed protocol and of the final protocol as a set of
equations. The composition process starts out from the initial
protocol equations and tries to reach the properties modeled by the
final equations. By doing so, they also prove the correctness of the
final protocol. In case of this approach, the human factor plays an
important role. As opposed to this, our approach can be fully
automatized, eliminating the interference of the human factor.

The approach proposed by S. Andova et all \cite{ACGMMR08} also uses
equations written for each protocol and for each security property
that must be satisfied by the final protocol. The composition
process uses the human operator to construct the final properties
from the initial equations and the Scyther \cite{CRE06b} tool to
automatically verify the correctness of the composed protocols. This
approach is a semi-automatized one that uses the human operator to
construct the final properties and an automatic verification tool
for the verification of the correctness of the final protocol.

\section{Conclusion and future work}\label{SEC:CONCLUSION}

We have developed a method for the composition of security
protocols. The novelty of our approach is the fact that it provides
a syntactical verification of the involved protocols, that makes it
appropriate for on-line automated composition applications.

Our proposal makes use of an enriched protocol model that embodies
protocol preconditions and effects. Messages exchanged by
participants are modeled as sequences of nodes called participant
chains. Based on this model we proposed conditions for the
precondition-effect composition. This process involves determining
if sufficient knowledge is provided by previous protocols and if
instance-specific security properties are maintained even after the
composition.

The protocol-chain composition process makes use of a canonical
model that eliminates message component instances. This model
reduces each component of the protocol model to its basic type. By
doing so we are able to verify the instance-independent components
of security protocols and detect multi-protocol attacks in a
syntactical manner.

We have applied the proposed composition method on several pairs of
well-known security protocols and have found new multi-protocol
attacks. Our independence verification method has been validated
using the security protocol verification tool Scyther, a state-space
exploration method, by discovering the same multi-protocol attacks.

As future work, we intend to use the proposed composition method in
the design process of new protocols for Web services. This would
allow us to implement more complex protocols, such as TLS
\cite{TLS99}, currently used as a binary security protocol, using an
XML message format that would enrich the properties of TLS with the
ones specific to Web services such as extensibility or flexibility.


\end{document}